\documentclass[final,5p,times]{elsarticle}
\usepackage{amsmath,amssymb}
\usepackage{xcolor}
\usepackage{fdsymbol}
\usepackage[T1]{fontenc}

\journal{Physics Letters A}

\begin{document}

\begin{frontmatter}
  \title{Steady-state phonon heat currents and differential thermal
    conductance across a junction of two harmonic phonon reservoirs}
  \author[iop]{Eduardo C. Cuansing}
  \ead{eccuansing@up.edu.ph}
  \author[iop]{Juan Rafael K. Bautista}
  \ead{jkbaustista@up.edu.ph}
  \affiliation[iop]{organization={Institute of Physics, University of the
      Philippines Los Ba\~{n}os},
    state={Laguna},
    postcode={4031},
    country={Philippines}}
  
  \begin{abstract}
    We study phonon transport in junctions of two harmonic reservoirs coupled
    together by a spring. The exact steady-state heat currents and thermal
    conductance are calculated using nonequilibrium Green's functions. We
    find that the heat currents follow Fourier's law and the thermal
    conductance has a peak whenever the phonon spectra match. At lower
    temperatures, however, the thermal conductance maximum may not coincide
    with the spectra-matching peak due to the exclusion of higher-frequency
    phonons, whose spectra may match, from participating in the transport.
    Furthermore, we find that increasing the coupling spring constant
    increases the thermal conductance. Lastly, the magnitude of the
    steady-state heat currents and thermal conductance are the same whether
    the direction of phonon flow is from left to right or vice versa, even
    with mass and spring constant asymmetry. The properties of this basic
    model can serve as a reference for more complicated setups of phonon
    transport in molecular junctions.
  \end{abstract}
  

\end{frontmatter}

\section{Introduction}
\label{sec:intro}

The study of quantum transport in molecular and atomic junctions is an active
field of research due to the conceptual challenges of building an extensive
understanding of the properties of the system and the immense potential for
many practical applications in electronics, computing, and the search for
novel devices \cite{aradhya2013,gehring2019}. Research in the transport of
heat across molecular and atomic junctions
\cite{galperin2007a,dubi2011,mosso2019,cui2019,zhang2024}, in particular,
leads to quantum thermal circuits where components such as
thermal resistors \cite{song2016}, thermal capacitors \cite{portugal2021},
thermal inductors \cite{schilling2019}, thermal diodes
\cite{chang2006,wang2017,li2004,wu2009,kalantar2021}, thermal transistors
\cite{li2006,zou2026,joulain2016,malavazi2024}, thermal memory
\cite{kubytskyi2014,ordonez2019}, thermal switches
\cite{cuansing2010a,dutta2017,liu2024} and thermal heat valves
\cite{dutta2020,ronzani2018} operate at the molecular level in which
quantum-mechanical and nonequilibrium effects are relevant. These thermal
circuits could act as complimentary components to miniaturized electronic
devices or as stand-alone devices functioning via the flow of heat rather
than electricity. The heat flowing in a nanoscale system is mainly due to
the energy carried by its constituent particles, such as electrons, phonons,
photons, and spins \cite{rego1998,segal2003,dhar2008}. In phononics
\cite{cahill2003,li2012,maldovan2013,cahill2014}, the carriers of energy
are purely phonons which, in contrast to electrons whose flow can be
managed by an electromagnetic field, may be directed by temperature
gradients, mass gradings, spring constant variations, and anharmonic
phonon-phonon interactions. A quantum thermal circuit containing
all-phononic devices can be operated by attaching two heat reservoirs at
different temperatures and letting heat flow from the hot to the cold
reservoir. In this work, we investigate the thermal properties, i.e., the
heat currents and the differential thermal conductance, of a junction of
two harmonic phonon reservoirs in direct physical contact.
Section~\ref{sec:model} of this paper describes the model of the junction
of the two reservoirs. In Sec.~\ref{sec:observables}, we show how the
heat currents and the differential thermal conductance are calculated
using nonequilibrium Green's functions. Numerical results are shown and
discussed in Sec.~\ref{sec:results}. Lastly, Sec.~\ref{sec:summary} is
the summary and conclusion.

\section{Model}
\label{sec:model}

\begin{figure}[!h]
  \centering
  \includegraphics[width=0.8\columnwidth,clip]{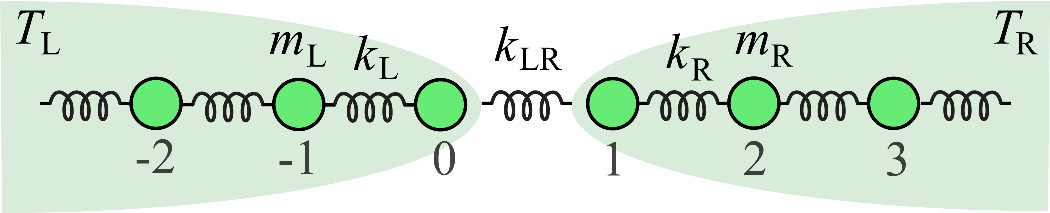}
  \caption{An illustration of two harmonic phonon reservoirs in contact
    via a coupling spring $k_{\mathrm{LR}}$. The left and right reservoirs
    have temperatures $T_{\mathrm{L}}$ and $T_{\mathrm{R}}$, site masses
    $m_{\mathrm{L}}$ and $m_{\mathrm{R}}$, and spring constants $k_{\mathrm{L}}$
    and $k_{\mathrm{R}}$, respectively.}
  \label{fig01}
\end{figure}

We consider the transport of heat in a junction made up of two heat
reservoirs that are electrical insulators but are sources of phonons.
Phonons \cite{srivastava2023}, therefore, are the sole carriers of energy
in the system. The direction of phonon flow is dictated by the temperature
difference between the two phonon reservoirs coupled together by a spring
$k_{\rm LR}$, as shown in Fig.~\ref{fig01}. Our model consists of three
parts, a left semi-infinite harmonic chain with temperature $T_{\rm L}$, a
right semi-infinite chain with temperature $T_{\rm R}$, and a central spring
that couples both chains. The semi-infinite chains serve as phonon
reservoirs and are described by the Hamiltonian:
\begin{align}
  \begin{split}
    H_{\rm L} & = \sum_{i=0}^{-\infty} \frac{p_i^2}{2 m_{\rm L}} + \frac{1}{2}
    \sum_{i=0}^{-\infty} k_{\rm L} \left(x_i - x_{i-1}\right)^2, \\
    H_{\rm R} & = \sum_{i=1}^{\infty} \frac{p_i^2}{2 m_{\rm R}} + \frac{1}{2}
    \sum_{i=1}^{\infty} k_{\rm R} \left(x_{i+1} - x_i\right)^2,
  \end{split}
  \label{eq:Hchains}
\end{align}
where $x_i$ and $p_i$ are the position and momentum of the oscillator at
site $i$, $m_{\rm L}$ and $m_{\rm R}$ are the masses and $k_{\rm L}$ and
$k_{\rm R}$ are the spring constants in the left and right chains,
respectively. The Hamiltonian corresponding to the coupling at the center is
\begin{align}
  H_{\rm LR} = \frac{1}{2} k_{\rm LR} \left( x_1 - x_0\right)^2,
  \label{eq:HLR}
\end{align}
where $k_{\rm LR}$ is the spring constant that couples the left and right
chains. The total Hamiltonian for the system is
$H = H_{\rm L} + H_{\rm R} + H_{\rm LR}$. In this model, we consider the two
reservoirs to be at their respective equilibrium at the start, i.e., when
$t \rightarrow -\infty$ with $H_0 = H_{\rm L} + H_{\rm R}$. The coupling
Hamiltonian $H_{\rm LR}$ is then switched on until the flow of phonons in
the system reaches its steady state.

\section{Heat currents and differential thermal conductance}
\label{sec:observables}

The rate of flow of heat out of the left reservoir, i.e., the left heat
current $J_{\rm L}(t)$, can be determined from the rate of change of the
energy in the left reservoir,
\begin{align}
  \begin{split}
    J_{\rm L}(t) \equiv -\left< \frac{d H_{\rm L}}{d t}\right>
    = -\frac{i}{\hbar} \Bigl\langle \bigl[H, H_{\rm L}\bigr]\Bigr\rangle.
  \end{split}
  \label{eq:JL1}
\end{align}
Noting that $[x_j,p_k] = i \hbar \delta_{jk}$, Eq.~(\ref{eq:JL1}) leads
to
\begin{equation}
  J_{\rm L}(t) = \hbar k_{\rm LR}~ {\rm Im}\!\left[-\frac{\partial
      G_{00}^<(t_1,t_2)}{\partial t_2} + \frac{\partial G_{10}^<(t_1,t_2)}
    {\partial t_2}\right]\Biggr|_t,
  \label{eq:JL}
\end{equation}
where we set $t_1=t_2=t$ after taking the derivatives, ${\rm Im}[]$ means
taking the imaginary part, and the lesser nonequilibrium Green's function is
$G_{jk}^<(t_1,t_2) = -\frac{i}{\hbar} \bigl\langle x_k(t_2)~ x_j(t_1)
\bigr\rangle$ where $j$ and $k$ are the labels of the sites in the chains.
Similarly, the heat current at the right reservoir is
\begin{equation}
  J_{\rm R}(t) = -\hbar k_{\rm LR}~ {\rm Im}\!\left[-\frac{\partial
      G_{11}^<(t_1,t_2)}{\partial t_2} + \frac{\partial G_{01}^<(t_1,t_2)}
    {\partial t_2}\right]\Biggr|_t,
  \label{eq:JR}
\end{equation}
where we also set $t_1=t_2=t$ after taking the derivatives. Because of the
way we define the heat currents, both $J_{\rm L}(t)$ and $J_{\rm R}(t)$ are
positive when the phonon flow is from the left to the right reservoir and
negative when the flow is in the reverse direction.

The transport of phonons in molecular junctions have previously been
studied utilizing the bilinear approximation
\cite{galperin2007a,galperin2007b}. In particular, a junction system of
phonon reservoirs, similar to the system that we consider in this work,
was previously studied employing the bilinear approximation to determine
the phonon current \cite{cuansing2010a}. In this work, in contrast, we
are not going to use the bilinear approximation. Instead, we consider the
full harmonic interaction for the coupling between the two reservoirs.
Define
\begin{equation}
  u = \left(\begin{array}{c}
    x_0 \\
    x_1 \\
  \end{array}\right)~~~~{\rm and}~~~~
  V = \frac{1}{2} k_{\rm LR} \left(\begin{array}{cc}
    1 & -1 \\
    -1 & 1 \\
  \end{array}\right).
  \label{eq:defuV}
\end{equation}
The coupling Hamiltonian in Eq.~(\ref{eq:HLR}), containing the full
harmonic interaction, can then be written as
\begin{equation}
  H_{\rm LR} = u^{\rm T} V u.
\end{equation}
We are going to use this form of $H_{\rm LR}$ in the perturbation expansion.
Defining the contour-ordered Green's function for sites $j$ and $k$ as
$G_{jk}(\tau_a,\tau_b) \equiv -\frac{i}{\hbar} \left\langle {\rm T}_c
\left[ x_j(\tau_a)\, x_k(\tau_b)\right]\right\rangle$ where ${\rm T}_c$ is
the contour-ordering operator and $\tau_a$ and $\tau_b$ are contour-time
variables along the Keldysh contour \cite{jauho1994,ridley2022,wang2014}.
Using the definitions in Eq.~(\ref{eq:defuV}), we can construct
\begin{align}
  \begin{split}
    G(\tau_a,\tau_b) & \equiv -\frac{i}{\hbar} \left\langle {\rm T}_c \left[
      u(\tau_a)\, u^{\rm T}(\tau_b)\right]\right\rangle \\
    & = \left(\begin{array}{cc}
    G_{00}(\tau_a,\tau_b) & G_{01}(\tau_a,\tau_b) \\
    G_{10}(\tau_a,\tau_b) & G_{11}(\tau_a,\tau_b) \\
  \end{array}\right).
  \end{split}
\end{align}
$G(\tau_a,\tau_b)$, therefore, contains the Green's functions that we need
to calculate the heat currents and the thermal conductance. We can consider
the Hamiltonian to be in two parts, i.e., $H=H_0+H_{\rm LR}$. The unperturbed
part is $H_0 = H_{\rm L} + H_{\rm R}$ while the perturbing part $H_{\rm LR}$
couples the two reservoirs. In the Interaction Picture, we therefore have
\begin{equation}
  G(\tau_a,\tau_b) = -\frac{i}{\hbar}\biggl\langle {\rm T}_c\left[
    e^{-\frac{i}{\hbar} \int_c H_{\rm LR}(\tau\rq)~d\tau\rq} u(\tau_a)\, u^{\rm T}(\tau_b)
    \right]\biggr\rangle_0,
  \label{eq:GIntPic}
\end{equation}
where the ensemble average is taken with respect to $H_0$. Expanding the
exponential and collecting terms, we end up with the Dyson equation
\begin{equation}
  G(\tau_1,\tau_b) = g(\tau_a,\tau_b) + \int_c g(\tau_a,\tau\rq)\,V(\tau\rq)\,
  G(\tau\rq,\tau_b)\,d\tau\rq
  \label{eq:Dyson}
\end{equation}
where $V$ is defined in Eq.~(\ref{eq:defuV}) and
\begin{align}
  \begin{split}
    g(\tau_a,\tau_b) & = -\frac{i}{\hbar} \biggl\langle {\rm T}_c\left[
      u(\tau_a)\,u^{\rm T}(\tau_b)\right]\biggr\rangle_0 \\
    & = \left(\begin{array}{cc}
      g_{00}(\tau_a,\tau_b) & 0 \\
      0 & g_{11}(\tau_a,\tau_b) \\
    \end{array}\right)
  \end{split}
\end{align}
contains the equilibrium Green's functions in the reservoirs. Using
analytic continuation and Langreth's theorem \cite{langreth1972,haug2008},
the retarded and advanced nonequilibrium Green's functions are
\begin{align}
  G^{\beta}(t_a,t_b) = g^{\beta}(t_a,t_b) + \int g^{\beta}(t_a,t\rq)\,V(t\rq)\,
  G^{\beta}(t\rq,t_b)\,dt\rq
  \label{eq:Gra}
\end{align}
where $\beta = {\rm r,a}$ and the lesser nonequilibrium Green's function is
\begin{align}
  \begin{split}
    G^<&(t_a,t_b) = g^<(t_a,t_b)
    + \int G^r(t_a,t\rq)\,V(t\rq)\,g^<(t\rq,t_b)\,dt\rq \\
    & + \int g^<(t_a,t\rq)\, V(t\rq)\, G^a(t\rq,t_b)\,dt\rq \\
    & + \int\!\int G^r(t_a,t')\,V(t')\,g^<(t',t'')\,V(t'')\,
    G^a(t'',t_b) dt'\,dt''
    \label{eq:Gless}
  \end{split}
\end{align}

The system satisfies time-translation invariance at the steady state. We
can, therefore, take the Fourier transforms of the Green's functions,
\begin{align}
  G^{\gamma}_{jk}(t_a,t_b) = \int_{-\infty}^{\infty} G^{\gamma}_{jk}(\omega)\,
  e^{-i\omega(t_a-t_b)}~\frac{d\omega}{2\pi},
\end{align}
where $\gamma = {\rm r,a,<}$. The frequency-domain steady-state heat
currents become
\begin{align}
  \begin{split}
    J_{\rm L}(\omega) & = \hbar\omega\, k_{\rm LR}\, {\rm Re}\!\left[
      G_{10}^<(\omega)-G_{00}^<(\omega)\right], \\
    J_{\rm R}(\omega) & = -\hbar\omega\, k_{\rm LR}\, {\rm Re}\!\left[
      G_{01}^<(\omega)-G_{11}^<(\omega)\right].
    \end{split}
  \label{eq:JLR}
\end{align}
The total heat current is calculated from
\begin{equation}
  J_{\alpha} \equiv \int_{-\infty}^{\infty} J_{\alpha}(\omega)~d\omega,
  \label{eq:totalcurrent}
\end{equation}
where $\alpha = {\rm L,R}$. The thermal conductance is then determined
from $K_{\alpha} \equiv d J_{\alpha}/dT_{\alpha}$. Taking the Fourier
transforms of the Green's functions in Eq.~(\ref{eq:Gra}), the retarded
and advanced steady-state Green's functions are
\begin{align}
  G^{\beta}(\omega) = \left(1 - g^{\beta}(\omega)\,V\right)^{-1}\, g^{\beta}(\omega)
  \label{eq:Gbeta}
\end{align}
where $\beta = {\rm r,a}$ and $1$ is a $2\times2$ unit matrix. Furthermore,
the steady-state lesser Green's function in Eq.~(\ref{eq:Gless}) becomes
\begin{align}
  \begin{split}
    G^<(\omega) =~& g^<(\omega) + G^r(\omega)\,V\,g^<(\omega)
    + g^<(\omega)\,V\,G^a(\omega) \\
    & + G^r(\omega)\,V\,g^<(\omega)\,V\,G^a(\omega)
  \end{split}
  \label{eq:Glessomega}
\end{align}

The steady-state heat currents in Eq.~(\ref{eq:JLR}) can be written in
Landauer-like formulas. Notice that the steady-state lesser Green's
function in Eq.~(\ref{eq:Glessomega}) can be written as
\begin{equation}
  G^<(\omega) = \left(1 + G^r(\omega)\,V\right)\,g^<(\omega)\,\left(1 +
  V\,G^a(\omega)\right).
  \label{eq:Gless1}
\end{equation}
From Eq.~(\ref{eq:Gbeta}), we can have
\begin{align}
  \begin{split}
    1 + G^r(\omega)\,V = & G^r(\omega)\,\left(g^r(\omega)\right)^{-1}, \\
    1 + V\,G^a(\omega) = & \left(g^a(\omega)\right)^{-1}\,G^a(\omega).
  \end{split}
  \label{eq:ginv}
\end{align}
Equation~(\ref{eq:Gless1}) can then be written as
\begin{equation}
  G^<(\omega) = G^r(\omega)\,\Sigma^<(\omega)\,G^a(\omega),
  \label{eq:Glesssigma}
\end{equation}
where
\begin{align}
  \begin{split}
    \Sigma^<(\omega) & = \left(g^r(\omega)\right)^{-1}\,g^<(\omega)\,
    \left(g^a(\omega)\right)^{-1}\\
    & = \left(\begin{array}{cc}
      -i\,f_{\rm L}\,\Gamma_{00} & 0 \\
      0 & -i\,f_{\rm R}\,\Gamma_{11} \\
    \end{array}\right),
  \end{split}
  \label{eq:Sigmaless}
\end{align}
where $f_{\rm L,R}=(\exp(\hbar\omega/k_{\rm B} T_{\rm L,R})-1)^{-1}$ are the
Bose-Einstein distribution functions in the reservoirs and $\Gamma_{00}$
and $\Gamma_{11}$ are the diagonal elements of
\begin{align}
  \begin{split}
    \Gamma(\omega) & \equiv i\,\left(\left(g^a(\omega)\right)^{-1}
    - \left(g^r(\omega)\right)^{-1}\right) \\
    & = i\,\left(\begin{array}{cc}
      \left(g^a_{00}\right)^{-1}-\left(g^r_{00}\right)^{-1} &
      0 \\
      0 & \left(g^a_{11}\right)^{-1}-\left(g^r_{11}\right)^{-1} \\
    \end{array}\right).
  \end{split}
\end{align}
From the elements of the $G^<(\omega)$ matrix in Eq.~(\ref{eq:Glesssigma}),
the total heat current can be written in a Landauer-like form as
\begin{align}
  J_{\rm L} = \int_{-\infty}^{\infty}\,\hbar \omega\, T(\omega)\,
  \left(f_{\rm L}-f_{\rm R}\right)\,d\omega,
  \label{eq:landauerlike}
\end{align}
where $T(\omega) = k_{\rm LR}\,{\rm Im}\!\left[G^r_{10}(\omega)\,\Gamma_{00}\,
  G^a_{00}(\omega)\right]$ and the corresponding differential thermal
conductance is
\begin{align}
  K_{\rm L} = \int_{-\infty}^{\infty} \frac{\hbar^2 \omega^2}
  {k_{\rm B} T_{\rm L}^2}\, T(\omega)\,f_{\rm L}\left(f_{\rm L} + 1\right)\,
  d\omega.
  \label{eq:conductance}
\end{align}
Equation~(\ref{eq:conductance}) can be separated into two parts, a
temperature-independent part, $F_1 = \hbar\omega\,T(\omega)$, and a
temperature-dependent part, $F_2 = f_{\rm L} (f_{\rm L}+1)\, \hbar
\omega/k_{\rm B} T_{\rm L}^2$. The contributions of the masses and the spring
constants are contained in $F_1$. Lastly, from the properties of the
steady-state Green's functions we find that the heat currents and thermal
conductance are the same for the reservoirs, i.e.,
$J_{\rm L} = J_{\rm R} = J$ and $K_{\rm L} = K_{\rm R} = K$, respectively.

The equilibrium Green's functions in each reservoir can be determined from
its Hamiltonian, shown in Eq.~(\ref{eq:Hchains}), and the resulting
equations of motion \cite{cuansing2010b}. We get the retarded equilibrium
Green's function in the frequency domain as
\begin{equation}
  g_{jj}^r(\omega) = -\frac{1}{k_{\alpha}} \left\{-\frac{\Omega_1}{2k_{\alpha}}
  \pm \frac{1}{2k_{\alpha}}\sqrt{\Omega_2^2 - 4 k_{\alpha}^2}\right\},
\end{equation}
where $j$, the site label, is either $0$ or $1$ corresponding to $\alpha$
being either {\rm L} or ${\rm R}$, respectively, $\Omega_1 = m_{\alpha}
\left(\omega + i \eta\right)^2 - 2 k_{\alpha}$, $\Omega_2 = m_{\alpha}
\omega^2 - 2 k_{\alpha}$, and $\eta \rightarrow 0$. The advanced equilibrium
Green's function is
\begin{equation}
  g_{jj}^a(\omega) = \left(g_{jj}^r(\omega)\right)^{\ast}
\end{equation}
and the lesser equilibrium Green's function is
\begin{equation}
  g_{jj}^<(\omega) = 2 i f_{\alpha}\,{\rm Im}\!\left[
    g_{jj}^r(\omega)\right],
\end{equation}
where $f_{\alpha}$ is the Bose-Einstein distribution function of the $\alpha$
reservoir.

\section{Numerical results}
\label{sec:results}

In this work, the temperatures of the left and right reservoirs are
$T_{\rm L} = T_m + \Delta T/2$ and $T_{\rm R} = T_m - \Delta T/2$, respectively,
where $T_m$ is the midtemperature and $\Delta T$ is the temperature
difference. The thermal conductance $K_{\alpha}$ is calculated with respect
to the midtemperature $T_m$. Furthermore, we find that in the steady state
the behavior of the heat currents are exactly the same for the left and the
right reservoirs, i.e., the continuity equation is satisfied. The thermal
conductance, therefore, is also the same for both reservoirs.

There are several open parameters that can be varied in exploring the
properties of the junction. These are the midtemperature $T_m$, the
temperature difference $\Delta T$, the masses $m_{\rm L}$ and $m_{\rm R}$ of
the particles in the reservoirs, the spring constants $k_{\rm L}$ and
$k_{\rm R}$ between particles in the reservoirs, and the coupling constant
$k_{\rm LR}$ that couples both reservoirs. We vary each parameter
separately.

\begin{figure}[!h]
  \centering
  \input{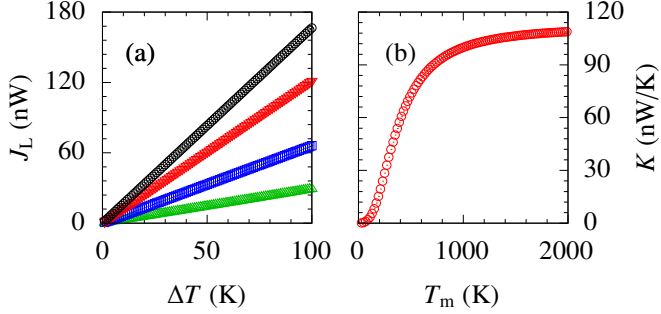}
  \vspace*{0.4in}
  \caption{(a) Plots of the heat currents $J_{\rm L}$ of the left
    reservoir as a function of the temperature difference $\Delta T$
    between the left and right reservoirs. The values of $T_{\rm m}$
    are $200$~K (green $\triangle$), $300$~K (blue $\Box$), $500$~K (red
    $\triangledown$), and $1000$~K (black $\medcircle$). (b) Plot of the
    thermal conductance $K$ as a function of the midtemperature
    $T_{\rm m}$.}
  \label{fig02}
\end{figure}

Shown in Fig.~\ref{fig02}(a) are plots of $J_{\rm L}$ as $\Delta T$
is varied for four different values of $T_m$. We find that for each $T_m$,
the heat current $J_{\rm L} \propto \Delta T$ linearly, following Fourier's law
for classical systems \cite{hahn2012}. Shown in Fig.~\ref{fig02}(b) is the
plot of the thermal conductance $K$ as a function of $T_m$. The plot for
thermal conductance is exactly the same for $\Delta T$ values ranging from
$10$~K to $40$~K and is the same as the plot calculated using the Landauer
formula for phonon transport across an infinitely-long linear chain with a
unit transmission coefficient \cite{wang2007}.

A positive $\Delta T$ means that the temperature of the left reservoir is
higher than that of the right reservoir. Phonons would therefore flow from
the left to the right. Inverting the sign of $\Delta T$ so that phonons
flow in the reverse direction, we find the same plots for the heat current
$J_{\rm L}$ and the thermal conductance $K$, as shown in
Fig.~\ref{fig02}, except for a change in sign because of the reversed
direction of phonon flow. 

\begin{figure}[!h]
  \centering
  \input{fig03a.tex}~~
  \input{fig03b.tex}
  \vspace*{0.3in}
  \caption{(a) Plot of $K$ as $k_{\mathrm{R}}$ and $T_{\mathrm{m}}$ are varied
    while maintaining $k_{\rm L}=k_{\rm LR}=1$~eV/\AA$^2$. (b) Plots of $K$ as
    $k_{\mathrm{R}}$ is varied for $T_{\mathrm{m}}$ values $2000$~K (green
    $\triangle$), $1000$~K (blue $\Box$), $500$~K (red $\triangledown$),
    and $300$~K (black $\medcircle$). (c) Plots of $K$ for lower
    $T_{\mathrm{m}}$ values: $150$~K (green $\triangle$), $200$~K (blue
    $\Box$), $250$~K (red $\triangledown$), and $300$~K (black $\medcircle$).
    The dash lines in (b) and (c) show when
    $k_{\mathrm{R}} = k_{\mathrm{L}} = 1$~eV/\AA$^2$.}
  \label{fig03}
\end{figure}

\begin{figure}[!h]
  \centering
  \includegraphics[width=0.45\textwidth,clip]{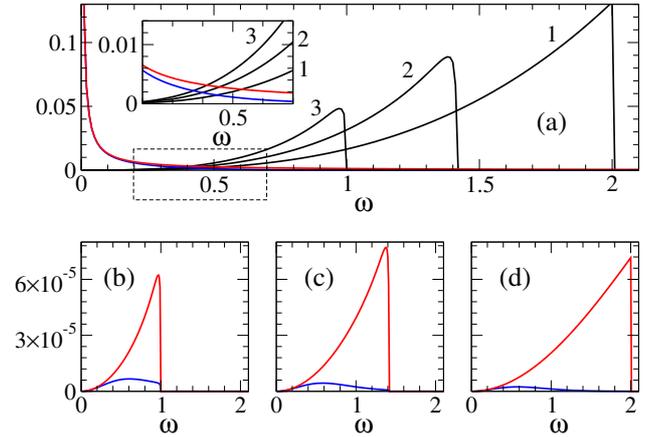}
  \vspace*{-0.1in}
  \caption{Plots of the terms in the thermal conductance, see
    Eq.~(\ref{eq:conductance}), for a range of frequencies $\omega$.
    (a) The temperature-independent $F_1$ terms are black solid lines
    with $k_{\rm R} = 1$~eV/\AA$^2$ for curve 1, $k_{\rm R} = 0.5$~eV/\AA$^2$
    for curve 2, and $k_{\rm R} = 0.25$~eV/\AA$^2$ for curve 3. The
    temperature-dependent $F_2$ terms are shown for $T_{\rm m} = 100$~K
    (blue solid line) and $T_{\rm m} = 1000$~K (red solid line). The inset
    shows an enlarged portion of the plot that is enclosed by the dashed
    rectangle. The products of $F_1$ and $F_2$ are shown in (b) for
    $k_{\rm R} = 0.25$~eV/\AA$^2$, (c) for $k_{\rm R} = 0.5$~eV/\AA$^2$, and
    (d) for $k_{\rm R} = 1$~eV/\AA$^2$, for temperatures $T_{\rm m} = 100$~K
    (blue solid line) and $T_{\rm m} = 1000$~K (red solid line). In all
    cases, $k_{\rm L} = k_{\rm LR} = 1$~eV/\AA$^2$, $m_{\rm L} = m_{\rm R} = 1$~amu,
    and $\Delta T = 30$~K.}
  \label{fig04}
\end{figure}

Shown in Fig.~\ref{fig03} are plots of $K$ when the spring constants in the
reservoirs are varied. The masses are maintained at
$m_{\rm L} = m_{\rm R} = 1$~amu and the midtemperature $T_m$ is varied from
$30$~K to $2000$~K. We set $k_{\rm L}=k_{\rm LR}=1$~eV/\AA$^2$ and vary
$k_{\rm R}$ from $0.1$~eV/\AA$^2$ to $2$~eV/\AA$^2$. A peak in $K$ appears at
$k_{\rm R}=1$~eV/\AA$^2$ when the phonon spectra of the reservoirs match.
However, at midtemperatures $T_m \leq 300$~K, the spectra-matching peak is
not the maximum of $K$. A soft maximum at $k_{\rm R} < 1$~eV/\AA$^2$ appears,
as can be seen in Fig.~\ref{fig03}(c). The reason for this maximum can be
deduced from the Landauer-like formula for the thermal conductance shown in
Eq.~(\ref{eq:conductance}). There are two competing terms in the formula for
the conductance, the temperature-independent $F_1$ term and the
temperature-dependent $F_2$ term. Contributions from the masses and the
spring constants are contained in $F_1$. Shown in Fig.~\ref{fig04} are
plots of $F_1$, $F_2$, and their products for a range of $\omega$. In
Fig.~\ref{fig04}(a), we see how $F_1$ varies with $k_{\rm R}$ and how $F_2$
varies with $T_{\rm m}$. For $F_1$, the largest enclosed area occurs when the
spectra matches at $k_{\rm R} = 1$~eV/\AA$^2$, i.e., see curve 1 of
Fig.~\ref{fig04}(a). Meanwhile, $F_2$ drops off at earlier frequencies when
the temperature is lower. The products of $F_1$ and $F_2$ are shown in
Fig.~\ref{fig04}(b) to (d) for various values of $k_{\rm R}$. From these
plots, we see that at higher temperatures, such as at $T_{\rm m} = 1000$~K,
the product between $F_1$ and $F_2$ includes the higher values of $F_1$ at
higher frequencies because of the long tail of $F_2$ at this temperature,
with a maximum value when the spectra matches. In contrast, at lower
temperatures, such as at $T_{\rm m} = 100$~K, $F_2$ drops off earlier at
lower frequencies thereby excluding the higher values of $F_1$ at higher
frequencies. The maximum of the product between $F_1$ and $F_2$, therefore,
does not occur when the spectra matches for such temperatures. In addition,
we find that inverting the sign of $\Delta T$ produces the same plots of
$K$ except for a change in sign because of the reversed direction of phonon
flow.

\begin{figure}[!h]
  \centering
  \input{fig05a.tex}~~
  \input{fig05b.tex}
  \vspace*{0.3in}
  \caption{(a) Plot of the thermal conductance $K$ as $T_{\mathrm{m}}$ and
    $m_{\mathrm{R}}$ are varied while maintaining $m_{\mathrm{L}}=1$~amu.
    (b) Plots of $K$ as $m_{\mathrm{R}}$ is varied for $T_{\mathrm{m}}$ values
    $2000$~K (green $\triangle$), $1000$~K (blue $\Box$), $500$~K (red
    $\triangledown$), and $300$~K (black $\medcircle$). In (c), the values
    of $T_{\mathrm{m}}$ are $150$~K (green $\triangle$), $200$~K (blue $\Box$),
    $250$~K (red $\triangledown$), and $300$~K (black $\medcircle$). The
    dash lines in (b) and (c) show when $m_{\mathrm{R}} = m_{\mathrm{L}} = 1$~
    amu.}
  \label{fig05}
\end{figure}

The effect of varying the masses of the particles to the thermal conductance
$K$ is shown in Fig.~\ref{fig05}. The spring constants are maintained at
$k_{\rm L}=k_{\rm R}=k_{\rm LR}=1$~eV/\AA$^2$. The midtemperature $T_m$ of the
reservoirs is varied from $30$~K to $2000$~K. We set $m_{\rm L} = 1$~amu in
the left reservoir and vary $m_{\rm R}$ from $0.1$~amu to $2$~amu. In
Fig.~\ref{fig05}, we see a peak in $K$ when $m_{\rm R} = 1$~amu when the
phonon spectra of the reservoirs match. At lower temperatures, such as at
$T_{\rm m} = 150$~K, however, the spectra-matching peak is not the maximum
of $K$. There is a soft maximum at a higher $m_{\rm R}$ value. This is
similar to the previous case when $k_{\rm R}$ was varied and a soft maximum
appears at a lower $k_{\rm R}$ value which does not correspond to spectra
matching. The only difference is that the soft maximum occurs at a higher
$m_{\rm R}$ value because of the reciprocal effects of $m_{\rm R}$ with
$k_{\rm R}$. We also inverted the sign of $\Delta T$ to test for any
asymmetry in phonon flow. We find that the magnitudes of the thermal
conductance $K$ are the same for either flow direction except for a change
in sign. There is no thermal rectification in this setup.

\begin{figure}[!h]
  \centering
  \input{fig06a.tex}~~
  \input{fig06b.tex}
  \vspace*{0.3in}
  \caption{(a) Plot of $K$ as $k_{\mathrm{LR}}$ and $T_{\mathrm{m}}$ are varied
    while holding $k_{\mathrm{L}} = k_{\mathrm{R}} = 1$~eV/\AA$^2$ and
    $m_{\rm L}=m_{\rm R}=1$~amu. (b) Plots of $K$ as $k_{\mathrm{LR}}$ is varied
    for $T_{\mathrm{m}}$ values $2000$~K (green $\triangle$), $1000$~K (blue
    $\Box$), $500$~K (red $\triangledown$), and $300$~K (black $\medcircle$).
    (c) Plots of $K$ for lower $T_{\mathrm{m}}$ values: $150$~K (green
    $\triangle$), $200$~K (blue $\Box$), $250$~K (red $\triangledown$), and
    $300$~K (black $\medcircle$). The dash lines in (b) and (c) show when
    $k_{\mathrm{LR}} = 1$~eV/\AA$^2$.}
  \label{fig06}
\end{figure}

The strength of the spring constant $k_{\rm LR}$ that couples the two
reservoirs can also be varied. Shown in Fig.~\ref{fig06} are plots of $K$
when $k_{\rm LR}$ is varied from $0.1$~eV/\AA$^2$ to $2$~eV/\AA$^2$. The
spring constants in the reservoirs are maintained at
$k_{\rm L}=k_{\rm R}=1$~eV/\AA$^2$ and the masses are also set at
$m_{\rm L}=m_{\rm R}=1$~amu. In Fig.~\ref{fig06}, the thermal conductance $K$
increases as the coupling $k_{\rm LR}$ and midtemperature $T_m$ are increased.
$K$ does not peak at $k_{\rm LR}=1$~eV/\AA$^2$. Instead, it keeps on
increasing as $k_{\rm LR}$ is increased allowing more phonons to flow across
the junction. Reversing the direction of phonon flow produces the same
plots of $K$ except for a change in sign.

\section{Summary and conclusion}
\label{sec:summary}

We have determined the phonon heat currents and thermal conductance in a
junction of two harmonic phonon reservoirs using nonequilibrium Green's
functions. The parameters that can be varied in the model include the
masses, the spring constants, the coupling spring constant, and the
temperatures of the reservoirs. Our model considers the full harmonic
interaction, as opposed to using the bilinear approximation, of the
coupling spring between the two reservoirs, with a coupling constant that
can have any value, from weak to strong coupling. Because the interactions
are all quadratic, we are able to find an exact Dyson equation for the
contour-ordered Green's function, leading to exact expressions for the
phonon heat currents and thermal conductance. Our results show that even
though our model and calculations are purely quantum-mechanical, the heat
currents satisfy the classical Fourier's law of heat conduction as the
difference in reservoir temperatures are varied. When the masses and the
spring constants in the reservoirs are varied, there are peaks in the thermal
conductance whenever the phonon spectra of the reservoirs match. The peaks,
however, may not be where the conductance is maximum. At low temperatures,
the higher-frequency phonons, where spectra matching can occur, are excluded
from participating in the transport. Furthermore, we find that the thermal
conductance increases when the strength of the spring that couples the two
reservoirs is increased. Finally, in the steady-state transport of phonons
across a junction of two harmonic phonon reservoirs, the magnitudes of the
phonon heat currents and thermal conductance are the same whether the flow
is from the left to the right reservoir or in the reverse direction, even
when the values of the masses or the spring constants are asymmetrical
between the two reservoirs. There is no thermal rectification. The
transport properties of this basic model of coupling two harmonic phonon
reservoirs can serve as a reference for further studies of phonon transport
in more complicated setups of molecular and atomic junctions.

\section*{Acknowledgments}

We are grateful to D.P.C. Dasallas for insightful discussions and to A.A.B.
Padama for giving us access to his group's computing facilities.

\bibliographystyle{elsarticle-num}
\bibliography{myreferences.bib}

@ARTICLE{aradhya2013,
AUTHOR  = {S. V. Aradhya and L. Venkataraman},
TITLE   = {Single-molecule junctions beyond electronic transport},
JOURNAL = {Nat.\ Nanotechnol.},
VOLUME  = {8},
PAGES   = {399},
YEAR    = {2013},
DOI     = {10.1038/nnano.2013.91}
}

@ARTICLE{gehring2019,
AUTHOR  = {P. Gehring and J. M. Thijssen and {H. S. J. van der Zant}},
TITLE   = {Single-molecule quantum-transport phenomena in break junctions},
JOURNAL = {Nat.\ Rev.\ Phys.},
VOLUME  = {1},
PAGES   = {381},
YEAR    = {2019},
DOI     = {10.1038/s42254-019-0055-1}
}

@ARTICLE{galperin2007a,
AUTHOR  = {M. Galperin and M. A. Ratner and A. Nitzan},
TITLE   = {Molecular transport junctions: vibrational effects},
JOURNAL = {J.\ Phys.:\ Condens.\ Matter},
VOLUME  = {19},
PAGES   = {103201},
YEAR    = {2007},
DOI     = {10.1088/0953-8984/19/10/103201}
}

@ARTICLE{dubi2011,
AUTHOR  = {Y. Dubi and {M. Di Ventra}},
TITLE   = {Colloquium: Heat flow and thermoelectricity in atomic and
           molecular junctions},
JOURNAL = {Rev.\ Mod.\ Phys.},
VOLUME  = {83},
PAGES   = {131},
YEAR    = {2011},
DOI     = {10.1103/RevModPhys.83.131}
}

@ARTICLE{mosso2019,
AUTHOR  = {N. Mosso and H. Sadeghi and A. Gemma and S. Sangtarash and
           U. Drechsler and C. Lambert and B. Gotsmann},
TITLE   = {Thermal transport through single-molecule junctions},
JOURNAL = {Nano\ Lett.},
VOLUME  = {19},
PAGES   = {7614},
YEAR    = {2019},
DOI     = {10.1021/acs.nanolett.9b02089}
}

@ARTICLE{cui2019,
AUTHOR  = {L. Cui and S. Hur and Z. A. Akbar and J. C. Kl\"{o}ckner and
           W. Jeong and F. Pauly and {S.-Y. Jang} and P. Reddy and
	   E. Meyhofer},
TITLE   = {Thermal conductance of single-molecule junctions},
JOURNAL = {Nature},
VOLUME  = {572},
PAGES   = {628},
YEAR    = {2019},
DOI     = {10.1038/s41586-019-1420-z}
}

@ARTICLE{zhang2024,
AUTHOR  = {H. Zhang and Y. Zhu and P. Duan and M. Shiri and S. Yelishala
           and S. Shen and Z. Song and C. Jia and X. Guo and L. Cui
	   and K. Wang},
TITLE   = {Energy conversion and transport in molecular-scale junctions},
JOURNAL = {App.\ Phys.\ Rev.},
VOLUME  = {11},
PAGES   = {041312},
YEAR    = {2024},
DOI     = {10.1063/5.0225756}
}

@ARTICLE{song2016,
AUTHOR  = {Q. Song and M. An and X. Chen and Z. Peng and J. Zang and
           N. Yang},
TITLE   = {Adjustable thermal resistor by reversibly folding a graphene
           sheet},
JOURNAL = {Nanoscale},
VOLUME  = {8},
PAGES   = {14943},
YEAR    = {2016},
DOI     = {10.1039/C6NR01992G}
}

@ARTICLE{portugal2021,
AUTHOR  = {P. Portugal and C. Flindt and {N. Lo Gullo}},
TITLE   = {Heat transport in a two-level system driven by a time-dependent
           temperature},
JOURNAL = {Phys.\ Rev.\ B},
VOLUME  = {104},
PAGES   = {205420},
YEAR    = {2021},
DOI     = {10.1103/PhysRevB.104.205420}
}

@ARTICLE{schilling2019,
AUTHOR  = {A. Schilling and X. Zhang and O. Bossen},
TITLE   = {Heat flowing from cold to hot without external intervention
           by using a "thermal inductor"},
JOURNAL = {Sci.\ Adv.},
VOLUME  = {5},
PAGES   = {eaat9953},
YEAR    = {2019},
DOI     = {10.1126/sciadv.aat9953}
}

@ARTICLE{chang2006,
AUTHOR  = {C. Chang and D. Okawa and A. Majumdar and Z. Zettl},
TITLE   = {Solid-state thermal rectifier},
JOURNAL = {Science},
VOLUME  = {314},
PAGES   = {1121},
YEAR    = {2006},
DOI     = {10.1126/science.1132898}
}

@ARTICLE{wang2017,
AUTHOR  = {H. Wang and S. Hu and K. Takahashi and X. Zhang and H. Takamatsu
           and J. Chen},
TITLE   = {Experimental study of thermal rectification in suspended
           monolayer graphene},
JOURNAL = {Nat.\ Commun.},
VOLUME  = {8},
PAGES   = {15843},
YEAR    = {2017},
DOI     = {10.1038/ncomms15843}
}

@ARTICLE{li2004,
AUTHOR  = {B. Li and L. Wang and G. Casati},
TITLE   = {Thermal diode: Rectification of heat flux},
JOURNAL = {Phys.\ Rev.\ Lett.},
VOLUME  = {93},
PAGES   = {184301},
YEAR    = {2004},
DOI     = {10.1103/PhysRevLett.93.184301}
}

@ARTICLE{wu2009,
AUTHOR  = {{L.-A. Wu} and C. X. Yu and D. Segal},
TITLE   = {Nonlinear quantum heat transfer in hybrid structures: Sufficient
           conditions for thermal rectification},
JOURNAL = {Phys.\ Rev.\ E},
VOLUME  = {80},
PAGES   = {041103},
YEAR    = {2009},
DOI     = {10.1103/PhysRevE.80.041103}
}

@ARTICLE{kalantar2021,
AUTHOR  = {N. Kalantar and B. K. Agarwalla and D. Segal},
TITLE   = {Harmonic chains and the thermal diode effect},
JOURNAL = {Phys.\ Rev.\ E},
VOLUME  = {103},
PAGES   = {052130},
YEAR    = {2021},
DOI     = {10.1103/PhysRevE.103.052130}
}

@ARTICLE{li2006,
AUTHOR  = {B. Li and L. Wang and G. Casati},
TITLE   = {Negative differential thermal resistance and thermal transistor},
JOURNAL = {Appl.\ Phys.\ Lett.},
VOLUME  = {88},
PAGES   = {143501},
YEAR    = {2006},
DOI     = {10.1063/1.2191730}
}

@ARTICLE{zou2026,
AUTHOR  = {Z. Zou and J. Gong and J. Wang and G. Casati and G. Benenti},
TITLE   = {Quantum vs classical thermal transport at low temperatures},
JOURNAL = {Phys.\ Rev.\ Lett.},
VOLUME  = {136},
PAGES   = {066305},
YEAR    = {2026},
DOI     = {10.1103/PhysRevLett.136.066305}
}

@ARTICLE{joulain2016,
AUTHOR  = {K. Joulain and J. Drevillon and Y. Ezzahri and J. Ordonez-Miranda},
TITLE   = {Quantum thermal transistor},
JOURNAL = {Phys.\ Rev.\ Lett.},
VOLUME  = {116},
PAGES   = {200601},
YEAR    = {2016},
DOI     = {10.1103/PhysRevLett.116.200601}
}

@ARTICLE{malavazi2024,
AUTHOR  = {A. H. A. Malavazi and B. Ahmadi and P. Mazurek and A. Mandarino},
TITLE   = {Detuning effects for heat-current control in quantum thermal
           devices},
JOURNAL = {Phys.\ Rev.\ E},
VOLUME  = {109},
PAGES   = {064146},
YEAR    = {2024},
DOI     = {10.1103/PhysRevE.109.064146}
}

@ARTICLE{kubytskyi2014,
AUTHOR  = {V. Kubytskyi and {S.-A. Biehs} and {P. Ben-Abdallah}},
TITLE   = {Radiative bistability and thermal memory},
JOURNAL = {Phys.\ Rev.\ Lett.},
VOLUME  = {113},
PAGES   = {074301},
YEAR    = {2014},
DOI     = {10.1103/PhysRevLett.113.074301}
}

@ARTICLE{ordonez2019,
AUTHOR  = {J. Ordonez-Miranda and Y. Ezzahri and J. A. Tiburcio-Moreno
           and K. Joulain and J. Drevillon},
TITLE   = {Radiative thermal memristor},
JOURNAL = {Phys.\ Rev.\ Lett.},
VOLUME  = {123},
PAGES   = {025901},
YEAR    = {2019},
DOI     = {10.1103/PhysRevLett.123.025901}
}

@ARTICLE{cuansing2010a,
AUTHOR  = {E. C. Cuansing and {J.-S. Wang}},
TITLE   = {Transient behavior of heat transport in a thermal switch},
JOURNAL = {Phys.\ Rev.\ B},
VOLUME  = {81},
PAGES   = {052302},
YEAR    = {2010},
DOI     = {10.1103/PhysRevB.81.052302}
}

@ARTICLE{dutta2017,
AUTHOR  = {B. Dutta and J. T. Peltonen and D. S. Antonenko and M. Meschke
           and M. A. Skvortsov and B. Kubala and J. K\"{o}nig and
	   C. B. Winkelmann and H. Courtois and J. P. Pekola},
TITLE   = {Thermal conductance of a single-electron transistor},
JOURNAL = {Phys.\ Rev.\ Lett.},
VOLUME  = {119},
PAGES   = {077701},
YEAR    = {2017},
DOI     = {10.1103/PhysRevLett.119.077701}
}

@ARTICLE{liu2024,
AUTHOR  = {C. Liu and C. Wu and Y. Zhao and Z. Chen and {T.-L. Ren} and
           Y. Chen and G. Zhang},
TITLE   = {Actively and reversibly controlling thermal conductivity in
           solid materials},
JOURNAL = {Phys.\ Rep.},
VOLUME  = {1058},
PAGES   = {1},
YEAR    = {2024},
DOI     = {10.1016/j.physrep.2024.01.001}
}

@ARTICLE{dutta2020,
AUTHOR  = {B. Dutta and D. Majidi and N. W. Talarico and {N. Lo Gullo}
           and H. Courtois and C. B. Winkelmann},
TITLE   = {Single-quantum-dot heat valve},
JOURNAL = {Phys.\ Rev.\ Lett.},
VOLUME  = {125},
PAGES   = {237701},
YEAR    = {2020},
DOI     = {10.1103/PhysRevLett.125.237701}
}

@ARTICLE{ronzani2018,
AUTHOR  = {A. Ronzani and B. Karimi and J. Senior and {Y.-C. Chang} and
           J. T. Peltonen and C. Chen and J. P. Pekola},
TITLE   = {Tunable photonic heat transport in a quantum heat valve},
JOURNAL = {Nat.\ Phys.},
VOLUME  = {14},
PAGES   = {991},
YEAR    = {2018},
DOI     = {10.1038/s41567-018-0199-4}
}

@ARTICLE{rego1998,
AUTHOR  = {L. G. C. Rego and G. Kirczenow},
TITLE   = {Quantized thermal conductance of dielectric quantum wires},
JOURNAL = {Phys.\ Rev.\ Lett.},
VOLUME  = {81},
PAGES   = {232},
YEAR    = {1998},
DOI     = {10.1103/PhysRevLett.81.232}
}

@ARTICLE{segal2003,
AUTHOR  = {D. Segal and A. Nitzan and P. H\"{a}nggi},
TITLE   = {Thermal conductance through molecular wires},
JOURNAL = {J.\ Chem.\ Phys.},
VOLUME  = {119},
PAGES   = {6840},
YEAR    = {2003},
DOI     = {10.1063/1.1603211}
}

@ARTICLE{dhar2008,
AUTHOR  = {A. Dhar},
TITLE   = {Heat transport in low-dimensional systems},
JOURNAL = {Adv.\ Phys.},
VOLUME  = {57},
PAGES   = {457},
YEAR    = {2008},
DOI     = {10.1080/00018730802538522}
}

@ARTICLE{cahill2003,
AUTHOR  = {D. G. Cahill and W. K. Ford and K. E. Goodson and G. D Mahan
           and A. Majumdar and H. J. Maris and R. Merlin and S. R. Phillpot},
TITLE   = {Nanoscale thermal transport},
JOURNAL = {J.\ Appl.\ Phys.},
VOLUME  = {93},
PAGES   = {793},
YEAR    = {2003},
DOI     = {10.1063/1.1524305}
}

@ARTICLE{li2012,
AUTHOR  = {N. Li and J. Ren and L. Wang and G. Zhang and P. H\"{a}nggi
           and B. Li},
TITLE   = {Collquium: Phononics: Manipulating heat flow with electronic
           analogs and beyond},
JOURNAL = {Rev.\ Mod.\ Phys.},
VOLUME  = {84},
PAGES   = {1045},
YEAR    = {2012},
DOI     = {10.1103/RevModPhys.84.1045}
}

@ARTICLE{maldovan2013,
AUTHOR  = {M. Maldovan},
TITLE   = {Sound and heat revolutions in phononics},
JOURNAL = {Nature},
VOLUME  = {503},
PAGES   = {209},
YEAR    = {2013},
DOI     = {10.1038/nature12608}
}

@ARTICLE{cahill2014,
AUTHOR  = {D. G. Cahill and P. V. Braun and G. Chen and D. R. Clarke and
           S. Fan and K. E. Goodson and P. Keblinski and W. P. King and
	   G. D. Mahan and A. Majumdar and H. J. Maris and S. R. Phillpot
	   and E. Pop and L. Shi},
TITLE   = {Nanoscale thermal transport. {II.} 2003-2012},
JOURNAL = {Appl.\ Phys.\ Rev.},
VOLUME  = {1},
PAGES   = {011305},
YEAR    = {2014},
DOI     = {10.1063/1.4832615}
}

@BOOK{srivastava2023,
AUTHOR    = {G. P. Srivastava},
TITLE     = {The Physics of Phonons},
EDITION   = {2nd},
YEAR      = {2023},
PUBLISHER = {CRC Press}
}

@ARTICLE{galperin2007b,
AUTHOR  = {M. Galperin and A. Nitzan and M. A. Ratner},
TITLE   = {Heat conduction in molecular transport junctions},
JOURNAL = {Phys.\ Rev.\ B},
VOLUME  = {75},
PAGES   = {155312},
YEAR    = {2007},
DOI     = {10.1103/PhysRevB.75.155312}
}

@ARTICLE{jauho1994,
AUTHOR  = {{A.-P. Jauho} and N. S. Wingreen and Y. Meir},
TITLE   = {Time-dependent transport in interacting and noninteracting
           resonant-tunneling systems},
JOURNAL = {Phys.\ Rev.\ B},
VOLUME  = {50},
PAGES   = {5528},
YEAR    = {1994},
DOI     = {10.1103/PhysRevB.50.5528}
}

@ARTICLE{ridley2022,
AUTHOR  = {M. Ridley and N. W. Talarico and D. Karlsson and {N. Lo Gullo} and
R. Tuovinen},
TITLE   = {A many-body approach to transport in quantum systems: from the
           transient regime to the stationary state},
JOURNAL = {J.\ Phys.\ A: Math.\ Theor.},
VOLUME  = {55},
PAGES   = {273001},
YEAR    = {2022},
DOI     = {10.1088/1751-8121/ac7119}
}

@ARTICLE{wang2014,
AUTHOR  = {{J.-S. Wang} and B. K. Agarwalla and H. Li and J. Thingna},
TITLE   = {Nonequilibrium {Green's} function method for quantum thermal
           transport},
JOURNAL = {Front.\ Phys.},
VOLUME  = {9},
PAGES   = {673},
YEAR    = {2014},
DOI     = {10.1007/s11467-013-0340-x}
}

@ARTICLE{langreth1972,
AUTHOR  = {D. C. Langreth and J. W. Wilkins},
TITLE   = {Theory of spin resonance in dilute magnetic alloys},
JOURNAL = {Phys.\ Rev.\ B},
VOLUME  = {6},
PAGES   = {3189},
YEAR    = {1972},
DOI     = {10.1103/PhysRevB.6.3189}
}

@BOOK{haug2008,
AUTHOR    = {H. Haug and {A.-P. Jauho}},
TITLE     = {Quantum Kinetics in Transport and Optics of Semiconductors},
YEAR      = {2008},
PUBLISHER = {Springer, Berlin}
}

@ARTICLE{cuansing2010b,
AUTHOR  = {E. C. Cuansing and {J.-S. Wang}},
TITLE   = {Tunable heat pump by modulating the coupling to the leads},
JOURNAL = {Phys.\ Rev.\ E},
VOLUME  = {82},
PAGES   = {021116},
YEAR    = {2010},
DOI     = {10.1103/PhysRevE.82.021116}
}

@BOOK{hahn2012,
AUTHOR    = {D. W. Hahn and {M. Necati \"{O}zisik}},
TITLE     = {Heat Conduction},
EDITION   = {3rd},
YEAR      = {2012},
PUBLISHER = {Wiley}
}

@ARTICLE{wang2007,
AUTHOR  = {{J.-S. Wang} and N. Zeng and J. Wang and C. K. Gan},
TITLE   = {Nonequilibrium {Green's} function method for thermal transport
           in junctions},
JOURNAL = {Phys.\ Rev.\ E},
VOLUME  = {75},
PAGES   = {061128},
YEAR    = {2007},
DOI     = {10.1103/PhysRevE.75.061128}
}
  
\end{document}